\documentclass[prapplied,longbibliography,
superscriptaddress,
twocolumn,
groupedauthors]{revtex4-1} 
\usepackage{graphicx}
\usepackage{hyperref}
\usepackage{dcolumn}   
\usepackage{bm}        
\usepackage{amssymb}   
\usepackage[english]{babel} 
\usepackage{inputenc}
\usepackage{amsmath}
\usepackage{amssymb}
\usepackage{graphicx}
\usepackage{mathrsfs}
\usepackage{color}
\usepackage{upgreek}

\begin{document}
\title{Engineering the dissipation of crystalline micromechanical resonators}
\author{Erick Romero}
\email{e.romero@uq.edu.au}
\affiliation{Centre for Engineered Quantum Systems, School of Mathematics and Physics, The University of Queensland, Australia}
\author{Victor M. Valenzuela}%
\affiliation{Facultad de Ciencias 
F\'isico-Matem\'aticas, Universidad Aut\'onoma de Sinaloa, M\'exico}
\author{Atieh R. Kermany}
\affiliation{Centre for Engineered Quantum Systems, School of Mathematics and Physics, The University of Queensland, Australia}
\author{Leo Sementilli}
\affiliation{Centre for Engineered Quantum Systems, School of Mathematics and Physics, The University of Queensland, Australia}
\author{Francesca Iacopi}
\affiliation{School of Electrical and Data Engineering, 
University of Technology Sydney, NSW, Australia}
\author{Warwick P. Bowen}
\affiliation{Centre for Engineered Quantum Systems, School of Mathematics and Physics, The University of Queensland, Australia}

\begin{abstract}

High quality micro- and nano-mechanical resonators are widely used in sensing, communications and timing, and have future applications in quantum technologies and fundamental studies of quantum physics. Crystalline thin-films are particularly attractive for such resonators due to their prospects for high quality, intrinsic stress and yield strength, and low dissipation. However, when grown on a silicon substrate, interfacial defects arising from lattice mismatch with the substrate have been postulated to introduce additional dissipation. Here, we develop a new backside etching process for single crystal silicon carbide microresonators that allows us to quantitatively verify this prediction. By engineering the geometry of the resonators and removing the defective interfacial layer, we achieve quality factors exceeding a million in silicon carbide trampoline resonators at room temperature, a factor of five higher than without the removal of the interfacial defect layer. We predict that similar devices fabricated from ultrahigh purity silicon carbide and leveraging its high yield strength, could enable room temperature quality factors as high as $6\times10^9$.
\end{abstract}

\maketitle

\section{Introduction}

Micro- and nano-mechanical resonators have a wide range of applications in industry and fundamental science, ranging from precision sensing of mass~\cite{yang_zeptogram-scale_2006}, single-molecules~\cite{chien_single-molecule_2018},
~ultrasound~\cite{basiri-esfahani_precision_2019}, magnetic fields~\cite{li_quantum_2018} and inertia~\cite{fan_graphene_2019}; to tests of spontaneous collapse models in quantum mechanics~\cite{vinante_improved_2017,forstner_testing_2019}, and memories and interfaces for quantum computers~\cite{bagheri_dynamic_2011,jiang_efficient_2019}. Achieving a high resonator quality factor is critical for many of these applications. 

Recently, remarkable progress has been made in improving the quality factor of micro- and nano-mechanical resonators fabricated from highly stressed thin amorphous films -- most particularly amorphous silicon nitride -- on a silicon substrate~\cite{reinhardt_ultralow-noise_2016,norte_mechanical_2016,tsaturyan_ultracoherent_2017,rossi_measurement-based_2018,ghadimi_radiation_2017,ghadimi_elastic_2018}. This progress has been achieved through a combination of dissipation engineering~\cite{tsaturyan_ultracoherent_2017}, to decrease both external energy loss to the environment and internal material losses, and strain engineering to approach the material yield strength and thereby dilute the dissipation~\cite{ghadimi_elastic_2018,fedorov_generalized_2019}. However, these strategies are now approaching their limits for amorphous materials. 

Crystalline materials offer a range of advantages that could allow them to go beyond these limits. High purity crystalline materials have a lower density of defects than amorphous materials, allowing significantly higher intrinsic quality factors. For instance, intrinsic quality factors above $10^5$ have been reported for highly pure diamond~\cite{najar_high_2014,tao_single-crystal_2014}, calcium fluoride~\cite{hofer_cavity_2010} and silicon carbide~\cite{jiang_sic_2019}, which has exhibited quality factors higher than 10$^6$ when surface losses have been eliminated~\cite{hamelin_monocrystalline_2019}. This compares to 25,000 in amorphous silicon nitride~\cite{villanueva_evidence_2014} and 1000 in amorphous silicon~\cite{gaspar_amorphous_2004}. Furthermore, due to crystal lattice mismatch, crystalline materials can be grown with high intrinsic stress~\cite{buckle_stress_2018}, crucial for dissipation dilution. Strained single crystal string resonators have been reported with quality factors exceeding 10$^5$ for GaNAs and 10$^6$ for SiC~\cite{onomitsu_ultrahigh-_2013,kermany_microresonators_2014}, while GaAs and In$_x$Ga$_{1-x}$P nanomembranes have reached quality factors above $10^6$~\cite{liu_high-q_2011,cole_tensile-strained_2014}. Moreover, crystalline materials have a high yield strength, increasing their potential for applications using both dissipation and strain engineering. For instance, crystalline silicon carbide thin films can exhibit intrinsic stress as high as 1.5~GPa~\cite{kermany_factors_2016} and have a yield strength of 21~GPa~\cite{petersen_silicon_1982}. This compares to 1.3~GPa and 6~GPa for amorphous silicon nitride. However, even with these significant advantages, thin-film crystalline resonators have not seen the same dramatic improvements in quality as their amorphous counterparts. This is due to in part increased complexity of fabrication~\cite{tao_single-crystal_2014,liu_high-q_2011,cole_tensile-strained_2014}, and -- when grown on a silicon substrate -- to the presence of dislocations and high-density of stacking faults in close proximity to the interface, which have been postulated to degrade the mechanical quality factor~\cite{la_via_thin_2018, anzalone_carbonization_2017, iacopi_evidence_2013,cleland_noise_2002}.

Here, we develop a new backside etching technique for the fabrication of crystalline thin-film resonators. We use this to fabricate high quality single-crystal silicon carbide trampoline resonators, and to characterise the effect of interfacial defect layer on their mechanical quality factor. By measuring  trampoline resonators of varying thicknesses, both back- and front-side etched, we are able to build a quantitative model to extract the volumetric intrinsic quality factors of the interfacial layer and of the high quality silicon carbide far from the interface, as well as the intrinsic surface quality factor. We find that defects degrade the volumetric quality factor by more than an order of magnitude near the interface, and achieve a diluted quality factor exceeding one million in devices for which the interfacial defect layer is removed. Our model predicts that diluted quality factors as high as $6\times10^9$ may be possible using both dissipation and strain engineering in high purity single crystal silicon carbide.

\section{Device Design and Fabrication}

\begin{figure}[h!]
\includegraphics[width=\columnwidth]{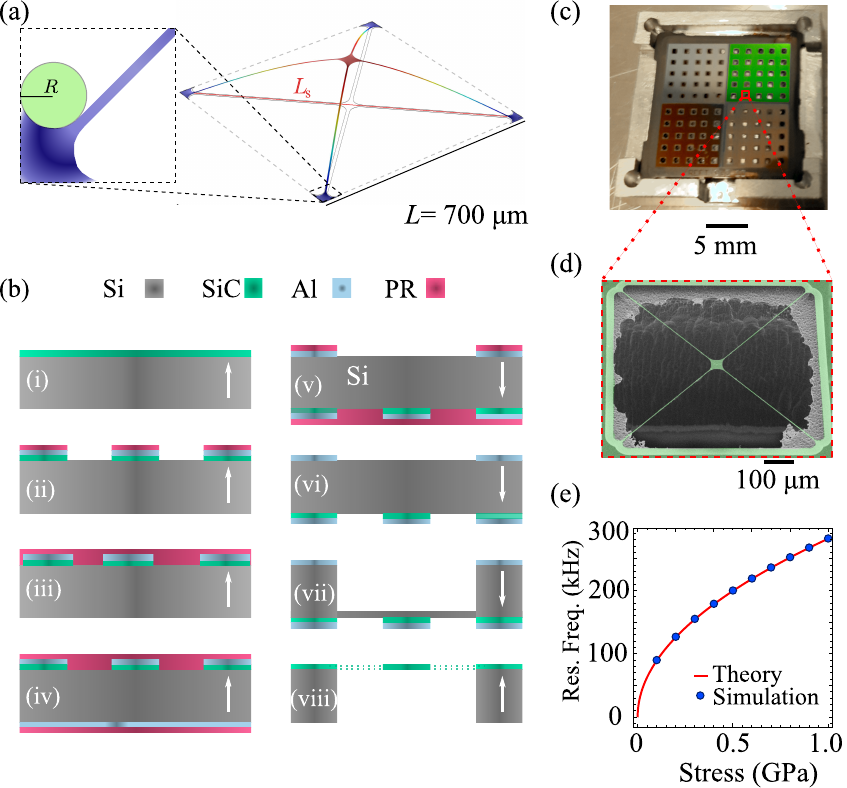}
\caption{
(a) Modeshape of the trampoline's fundamental vibrational out-of-plane mode with lateral length $L=700~\upmu$m and clamping points rounded with $R=30~\upmu$m (inset). (b) Fabrication steps for single crystal SiC trampoline resonators. (c) Image of a SiC on Si trampolines chip sitting on an aluminium holder. The different colors of the four different chip sections are produced by the different thickness of the SiC. Each device is released on a fully etched silicon window which can be directly observed in the image. (d) SEM image of a SiC trampoline resonator. The trampoline is suspended above a fully back side etched hole in a 0.5~mm thick of silicon substrate. (e) Resonance frequency of a trampoline calculated numerically using finite element modelling (blue dots) compared with the resonance frequency of a string of length $L_s=\sqrt{2}L$ calculated analytically (red line) using Eq.~\eqref{eq:Frequency} with thickness $h=337~$nm.}\label{fig:Fabrication}
\end{figure}

Our trampoline design features a square inner island of 40~$\upmu$m side suspended within a 700~$\upmu$m$\times$700~$\upmu$m squared hollow by four 5~$\upmu$m wide and $\sim 500~\upmu$m long tethers. These are connected to the substrate by adiabatically widened and rounded clamping points with radius of curvature $R$, as shown in the inset of Fig.~\ref{fig:Fabrication}(a). In contrast to clamp-tapered approaches, where the dissipation dilution is achieved through localized stress~\cite{bereyhi_clamp-tapering_2019}, the dissipation dilution present in our trampolines comes from clamping points with radius of curvature engineered to optimize the fraction of elastic energy stored as elongation as opposed to bending~\cite{schmid_fundamentals_2016,tsaturyan_ultracoherent_2017}. In this approach, two counteracting and competing mechanisms are present; the increased material volume at the widened clamps stores a larger amount of bending energy, while the increased rigidity reduces the overall bending~\cite{romero_sanchez_phononics:_2019,sadeghi_influence_2019}. The elongation to bending ratio converges to a maximum value for an optimum radius of curvature $R$. According to our finite element simulations, the optimal radius of curvature for our crystalline resonators is $R=30~\mu$m, predicting a $Q$ about four times larger compared to rigid-clamping ($R=0$).

The fabrication of crystalline trampoline resonators introduces technical hurdles that are not present during the fabrication of highly stressed amorphous trampolines~\cite{reinhardt_ultralow-noise_2016,norte_mechanical_2016}. One major difference is that, the thin-film growth of amorphous materials is independent of the substrate crystal orientation, while crystalline thin-films, which are often seeded by the crystal orientation of the substrate, being dependent of the crystal orientation. The crystal lattice mismatch between the thin-film and the seeding substrate induces a high density of crystal defects through the first few nanometers of the film, accessible by back-side etch. Release methods used for amorphous materials on silicon such as isotropic wet etch become ineffective for some crystalline films as the etch rate is strongly dependent on the silicon crystal orientation. For this reason, we develop an alternative back-side etch technique for the release of crystalline resonators compatible with all crystal orientations of the Si substrate.

Our trampoline resonators are fabricated from a highly-stressed 3C-SiC single crystal thin film developed at the Queensland Microtechnology Facility (QMF), within the Queensland node of the Australian National Fabrication Facility. The SiC material of initial thickness $h=337~$nm is grown by hetero-epitaxial deposition atop a 500~$\upmu$m thick Si substrate, as depicted in Fig.~\ref{fig:Fabrication}(b-i). Using standard photolithography, the trampolines are first patterned on an aluminium thin film (150~nm) which has been evaporated atop the SiC. The wet etched aluminium functions as a hard mask and as a protective layer during handling. The pattern is transferred to the SiC by Reactive Ion Etching (RIE), see {Fig.~\ref{fig:Fabrication}(b-ii)}. To protect the patterned aluminium and SiC during later manipulation, a layer of photoresist is spin coated as illustrated in {Fig.~\ref{fig:Fabrication}(b-iii)}. A 150~nm film of aluminium is evaporated on the back-side, which will be used as a hard mask during the back-side etch, and the aluminium film is protected with positive photoresist (Fig.~\ref{fig:Fabrication}(b-iv)). The front- and back-side patterns, with their corresponding alignment marks, are aligned with the front and rear optical microscope objectives of an EVG620 mask aligner. The resist is patterned defining squared windows which are transferred to the aluminium using wet etch (Fig.~\ref{fig:Fabrication}(b-v)). The wafer is placed facing down on a secondary carrier wafer, with a thin coating of fomblin oil between them to enhance thermal contact~(Fig.~\ref{fig:Fabrication}(b-vi)). The silicon is back-side deep-etched $\sim$480~$\upmu$m using Deep Rective Ion Etching (DRIE) as represented in Fig.~\ref{fig:Fabrication}(b-vii). The chip is then immersed in a heated (80$^{\circ}$~C) potassium hydroxide solution to remove the aluminium mask and the excess of silicon ``grass'' formed during the etching process~\cite{jung_parameter_2010}.

To elucidate the effect of the interface crystal defects of the released SiC structures on the mechanical $Q$, a selective front-side etch is used to vary the thickness $h$ of the SiC film. The front-side thinning is performed by masking the chip in sections and etching the SiC layer using RIE dry etch. An image of the chip after etch is presented in Fig.~\ref{fig:Fabrication}(c), where the evident change in color for different film thickness is produced by the thin-film interference effect of the illumination light. The trampolines are released using a XeF$_2$ chemical dry etch of silicon, see Fig.~\ref{fig:Fabrication}(b-viii). As an example, Fig.~\ref{fig:Fabrication}(d) shows a SEM image of a fully released trampoline resonator. The layer with high density of crystal defects near the Si-SiC interface ($>50$~nm)~\cite{iacopi_orientation-dependent_2013} becomes accessible once the trampolines are fully released. The chip is flipped and placed on a secondary substrate and 77~nm of SiC from the interface is removed in some trampolines using RIE dry back-side etch, eliminating most of the defect-rich layer.

\section{Experimental results}\label{Sec:Experiment}

\begin{figure}[h]
\includegraphics[width=\columnwidth]{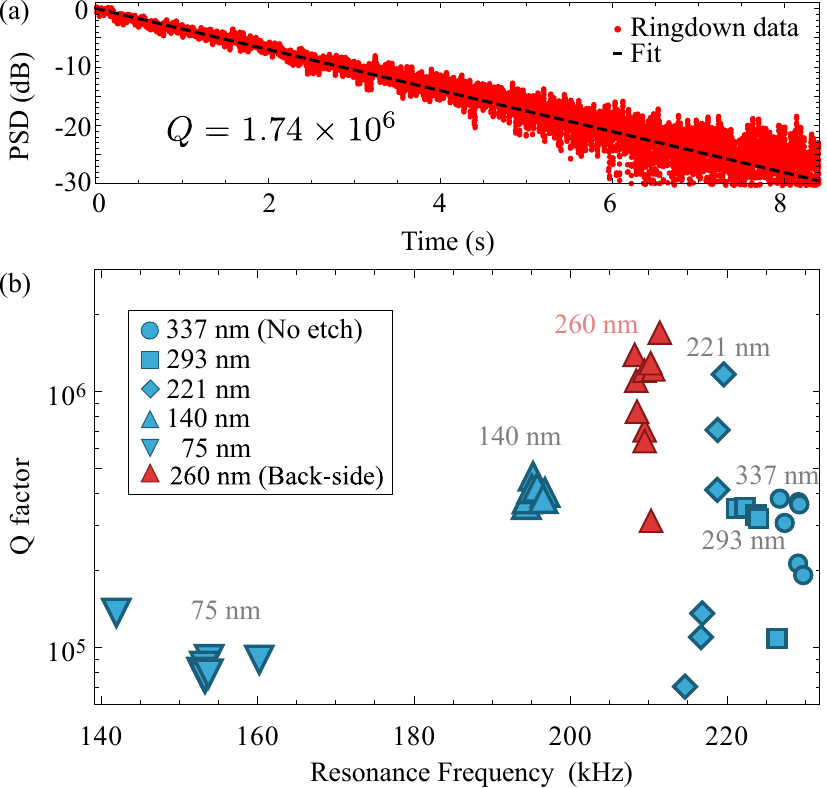}
\caption{(a) Normalized ringdown measurement on the fundamental mode of a trampoline resonator with resonance frequency $\omega_{\textnormal{m}}/2\pi\sim$211~kHz and mechanical quality factor $Q=1.7\times 10^6$. Ringdown fitting (black-dashed line) using linear regression of the free-ringdown signal captured in a single-shot. (b) Measured resonance frequencies $\omega_{\textnormal{m}}/2\pi$ and quality factors $Q$ of various single crystal SiC trampoline resonators. The blue color code  is for front-side etched devices. Each point represents an individual device and different shapes correspond to different thickness: circle $h=$337~nm, square $h=$293~nm, rhombus $h=$221~nm, triangle up $h=$140~nm and triangle down $h=$75~nm. The triangles in red represent the devices that were back-side etched with total thickness $h=$260~nm.  
}\label{fig:Ringdown}
\end{figure}

The resonance frequency of highly stressed resonators is strongly dependent on their internal mean stress $\sigma$ and scales as $\sqrt{\sigma}$. The non-uniform residual stress $\sigma_{\textnormal{r}}$ of hetero-epitaxially grown 3C-SiC~\cite{kermany_factors_2016} allows us to investigate the dependence of the mean stress on film thickness $\sigma(h)$. To find the relation $\sigma(h)$, we measure the fundamental resonance frequency $\omega_{\textnormal{m}}$ of each fabricated trampoline from its noise power spectral density with an  optical heterodyne detection system operating in vacuum ($P\sim 10^{-6}$~mbar) as reported previously~\cite{romero_propagation_2019,
romero_sanchez_phononics:_2019}. The quality factor $Q=\omega_{\textnormal{m}}/\Gamma$ is measured via ringdown after applying an impulse to the trampoline and measuring the power decay as $e^{-\Gamma t}$, with decay rate $\Gamma$. An example of ringdown measurement (red dots) is shown in Fig.~\ref{fig:Ringdown}(a) for the fundamental mode of a back-side etched trampoline of frequency $\omega_{\textnormal{m}}/2\pi=211~$kHz. The fit (black dashed line) is done using a linear regression of the free-ringdown signal, obtaining $\Gamma/2\pi \approx 1/(8.23~\textnormal{s})$. The measured $Q\approx 1.7\times 10^6$ of SiC trampolines is comparable to the $Q>10^6$ achieved in SiC strings~\cite{kermany_microresonators_2014}. Moreover, it compares favourably with the $Q \sim 10^5$ measured for GaNAs crystalline string resonators at room temperature~\cite{onomitsu_ultrahigh-_2013}, and is comparable to the $Q~\sim 2 \times 10^6$ of In$_x$Ga$_{1-x}$P and GaAs crystalline membrane resonators of similar dimensions ($L\sim 1$~mm)~\cite{cole_tensile-strained_2014,liu_high-q_2011}.

The fundamental resonance frequency $\omega_{\textnormal{m}}$ and quality factor $Q$ were measured on a total of 45 devices. These results are shown in Fig.~\ref{fig:Ringdown}(b) where each data point represents an individual device. Nine devices, identified as red triangles, were back-side etched by 77~nm, removing the defect-rich layer and leaving a final thickness of $h=260$~nm. Six devices with the original film thickness $h=337$~nm (circles) were measured without etching. The other 30 devices were front-side etched with final thicknesses $h=293$~nm (squares), $h=221$~nm (rhombuses), $h=140$~nm (triangles up) and $h=75$~nm (triangles down). Most of the measured devices have resonance frequencies $\omega_{\textnormal{m}}/2\pi \gtrsim 200~$kHz. However, devices that were largely front-side etched to a final thickness $h=75~$nm suffered from a significant decrease in both resonance frequency and quality factor. This dramatic decrease in $\omega_{\textnormal{m}}$ and $Q$ is caused by a substantial reduction of the mean stress $\sigma(h)$, and is consistent with observations that the layer near the interface is under compressive stress~\cite{iacopi_evidence_2013}.

\section{Model and Discussion}
 
The front- and back-side SiC etch is expected to have different effects on the $Q$ and resonance frequency of crystalline resonators due to the presence of crystal defects near the interface. In order to better understand and quantify the role these crystal defects play on the dissipation of crystalline resonators, we have developed a generic model for the total $Q$ that allows us to identify five different energy dissipation mechanisms grouped in two main categories, intrinsic and external. External energy dissipation is attributed to two main mechanisms, gas damping $Q_{\textnormal{gas}}^{-1}$ and clamping losses $Q_{\textnormal{clamp}}^{-1}$~\cite{schmid_fundamentals_2016}. Our resonators are characterized in a vacuum chamber that operates at $P\sim 10^{-6}$~mbar and room temperature, with a characteristic Knudsen number $K_n\sim 75, 000$, deeply within the molecular regime ($K_n >1$)~\cite{schmid_fundamentals_2016}. This rarefied gas environment allows us to neglect the damping caused by collisions between gas molecules and the resonator, leaving clamping losses as the main mechanism of external dissipation. These are detailed in Sec.~\ref{sec:Clamping}. 

\begin{figure}[h]
\includegraphics[width=\columnwidth]{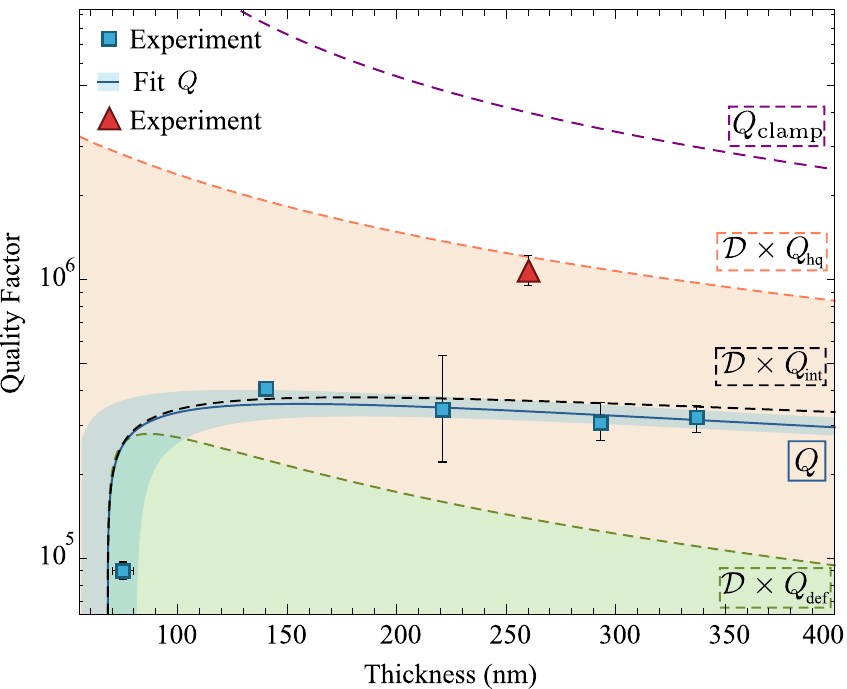}
\caption{Mean values of the measured $Q$ of SiC trampolines with different thickness $h$, obtained from the raw data presented in Fig. 2 (b). The blue squares (red triangles) are the front-side (back-side) thinned resonators and the error bars are the standard error. The blue line is the theoretical fit for $Q$ in Eq.~\eqref{eq:Q} for front-side etched resonators, the shaded region is the uncertainty of the fit. The mean $Q$ for the back-side etched resonators is shown as a red triangle. The green (orange) dashed line is $\mathcal{D}(h)\times Q_{\textnormal{def}}$ ($\mathcal{D}(h)\times Q_{\textnormal{hq}}$) the limit for trampoline resonators made solely from the SiC defect-rich (high-quality) layer. The purple dashed line is $Q_{\textnormal{clamp}}$ calculated from Eq.~\eqref{eq:clamping} for a SiC trampoline resonator with intrinsic stress $\sigma(h)$. The black dashed line is the diluted intrinsic quality factor $\mathcal{D}(h)\times Q_{\textnormal{int}}$, when clamping losses are neglected.}\label{fig:Results}
\end{figure}

Meanwhile, intrinsic dissipation $Q_{\textnormal{int}}^{-1}=Q_{\textnormal{vol}}^{-1}+Q_{\textnormal{surf}}^{-1}$ is caused by surface losses ($Q_{\textnormal{surf}}^{-1}$)~\cite{villanueva_evidence_2014} and intrinsic friction in the volume of the material ($Q_{\textnormal{vol}}^{-1}$)~\cite{ashby_overview_1989}, and is detailed in Sec.~\ref{sec:Intrinsic}. Thermoelastic losses in highly stressed trampoline resonators were calculated ($Q_{\textnormal{TED}}\sim 10^9$)~\cite{romero_sanchez_phononics:_2019} using existing models  \cite{lifshitz_thermoelastic_2000}, and are neglected in the rest of this work due to their small contributions in thin resonators~\cite{yasumura_quality_2000}. The total quality factor is given by 
\begin{equation}\label{eq:Q}
Q^{-1}=\mathcal{D}^{-1} Q_{\textnormal{int}}^{-1}+Q_{\textnormal{clamp}}^{-1},
\end{equation}
where $\mathcal{D}(h)$ is the dilution factor and is well approximated by the analytic expression for a string ${\mathcal{D}^{-1}(h)\approx (2 \lambda +\pi^2 \lambda^2)}$, where ${\lambda=(h/L)\sqrt{E/12 \sigma(h)}}$ and $L$ is the length of the string~\cite{schmid_fundamentals_2016}. In Fig.~\ref{fig:Results}, we plot the mean values of the experimentally measured $Q$ (blue squares) and the theoretical fit following Eq.~\eqref{eq:Q} (blue line). In the remainder of the paper we discuss in detail the model developed to fit the $Q$ as a function of thickness of a non-uniform stressed crystalline film.

\subsection{Stress Profile and Dissipation Dilution}\label{sec:Stress}

\begin{figure}[h]
\includegraphics[width=\columnwidth]{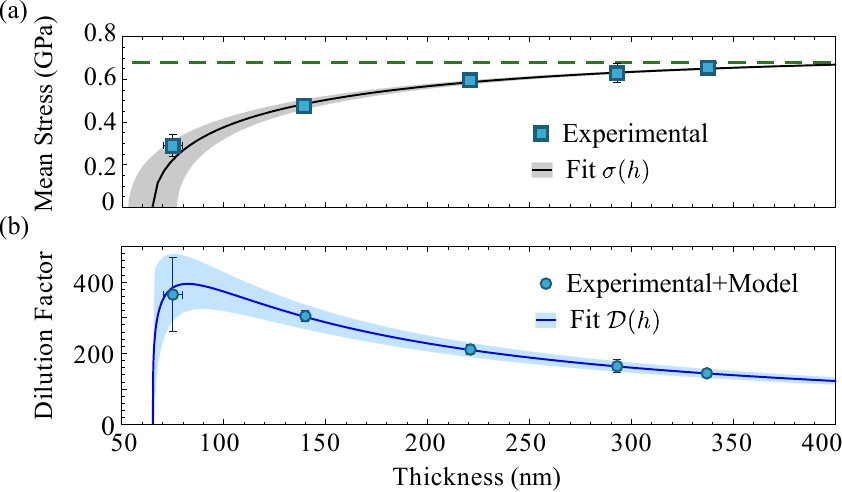}
\caption{(a) Mean stress $\sigma(h)$ as a function of the total thickness of the SiC thin film for front-side etched resonators. The experimental data are the mean values obtained using Eq.~\eqref{eq:StressThickness} from the measured resonance frequency. The error bars represent the standard error. The black line is the theoretical fit described in Eq.~\eqref{eq:StressThickness} and the gray band is the associated uncertainty of the fit. (b) Dilution factor $\mathcal{D}(h)$ for a thin film trampoline with mean stress $\sigma(h)$. The blue circles are determined from experimental data shown in Fig.~\ref{fig:Stress}(a). The light blue band is the uncertainty of the theoretical fit.}\label{fig:Stress}
\end{figure}

Quantifying $\mathcal{D}(h)$, $Q_{\textnormal{int}}$ and $Q_{\textnormal{clamp}}$ in Eq.~\eqref{eq:Q} requires us to identify the resonance frequency $\omega_{\textnormal{m}}(\sigma)$ and thickness dependent mean stress $\sigma(h)$. Front-side thinned resonators experience a reduction of the mean stress $\sigma(h)$, shown in Fig.~\ref{fig:Stress}(a) as blue squares. The monotonic decrement of stress as a function of $h$ is a direct consequence of the declining proportion of high stress material in the thinner structures. The supplier specifies the mean stress of the SiC film prior to any thinning process ($h=337$~nm) to be ${\sigma_{\textnormal{0}}\sim 620}$~MPa, shown in Fig~\ref{fig:Stress}(a) as a dashed line, which agrees reasonably with our measurements of ${\sigma_{\textnormal{0}}\sim 660}$~MPa. 

There is no known exact analytic solution for the resonance frequencies of a trampoline. However, we find that the fundamental resonance frequency can be accurately modeled by that of a string resonator of length $L_{\textnormal{s}} = \sqrt{2} L$, corresponding to the diagonal length of the trampoline. This is given by~\cite{schmid_fundamentals_2016},
\begin{equation}\label{eq:Frequency}
\omega_{\textnormal{m}}(\sigma)
=
\left(\frac{\pi}{2 L_s}\right)^2
\sqrt{\frac{ 4 E  h^2 }{3 \rho }}
\sqrt{1+ \frac{12  \sigma(h) L_s^2}{E  h^2 \pi^2}},
\end{equation}
where the density and Young's modulus of SiC are ${\rho =3210~\textnormal{kg/m}^3}$ and ${E=400}$~GPa, respectively~\cite{kermany_microresonators_2014}. To confirm the accuracy of this model, the analytic expression (red line) is compared to finite element modelling (blue dots) for the fundamental vibration mode of the trampoline resonators as a function of the intrinsic mean stress $\sigma$ of the SiC film. As shown in Fig.~\ref{fig:Fabrication}(e), we find very good agreement. 

To build an approximate analytical model of $\sigma(h)$ for front-side thinned resonators we attempt to fit it to several basic growth functional forms including the Logistic model~\cite{panik_growth_2014}, Gompertz's model~\cite{panik_growth_2014}, Bridges’s growth model~\cite{t._c._bridges_mathematical_1986} and Solow's model\cite{panik_growth_2014}. Of these, only a Bridges’s growth model~\cite{t._c._bridges_mathematical_1986} agrees reasonably with the mean stress. The final form of the model is
\begin{equation}\label{eq:StressThickness}
\sigma(h)= \displaystyle\sigma_{\textnormal{max}}\left(1- e^{- \left[c_1(h-h_0)\right]^{c_2}}\right),
\end{equation}
where $h_0=(68\pm 12)$~nm is the transition thickness at which the mean stress goes from compressive to tensile (\textit{i.e.} $\sigma(h_0)=0$). The exponential growth constant is $c_1=0.015~\textnormal{nm}^{-1}$
and $c_2=0.54$ is known as the kinetic order. Using this model, the predicted maximum stress of the high-quality SiC layer would be enhanced to ${\sigma_{\textnormal{max}}=740~\textnormal{MPa}}$ when there is no defect-rich layer.

From the mean stress $\sigma(h)$, it is also possible to estimate the functional form of the layer by layer residual stress $\sigma_{\textnormal{r}}(z)$. The residual stress is produced during the hetero-epitaxial growth of 3C-SiC thin-films and is related to the mean stress of the film through the relation $\sigma(h)=\frac{1}{h}\int_0^{h} \sigma_{\textnormal{r}}(z)~dz$. So far, its functional form has not been well known because existing spectroscopic methods to measure $\sigma_{\textnormal{r}}(z)$ are incompatible with thin film analysis. For example Raman spectroscopy suffers from limited resolution ($\sim1~\upmu$m)~\cite{anzalone_advanced_2011}, while other mesurement methods such as bulge testing, are incompatible with micro- and nano-mechanical systems as they are limited to centimeter scale sealed membranes~\cite{iacopi_evidence_2013}. Our method to characterize the mean stress for released microstructures could in future provide a precise determination of the residual stress profile $\sigma_{\textnormal{r}}(z)$ of hetero-epitaxial thin films beyond the resolution of existing non-invasive methods~\cite{anzalone_advanced_2011,iacopi_evidence_2013}.

The experimental $\mathcal{D}(h)$ shown in Fig.~\ref{fig:Stress}(b) (blue points) is determined by combining the measurements of the stress $\sigma(h)$ (see Fig.~\ref{fig:Stress}(a)) and the analytical model of the thickness dependent dilution factor $\mathcal{D}(h)$ for a string of length $L_s$~\cite{schmid_fundamentals_2016}. The good agreement provides experimental validation that it is appropriate to model trampolines as strings relevant both to our work and previous research~\cite{norte_mechanical_2016,sadeghi_influence_2019,bereyhi_clamp-tapering_2019}. The combined results from Fig.~\ref{fig:Stress} suggest that the highest enhancement to the quality factor should occur at a film thickness of $h\sim$70~nm. However, the total $Q$ is affected not just by the dilution factor but also by intrinsic loss mechanisms and clamping losses.

In highly stressed resonators, clamping losses are often neglected as the largest contribution to the loss comes from intrinsic mechanisms. In this limit, the intrinsic quality factor $Q_{\textnormal{int}}$ is estimated dividing $Q\approx \mathcal{D}(h)\times Q_{\textnormal{int}}$ (black dashed line in Fig.~\ref{fig:Results}), by the dilution factor $\mathcal{D}(h)$. However, for our trampolines we find the experimental results deviate from this model at large thicknesses as it is shown in Fig.~\ref{fig:Results} (blue line), indicating that clamping losses should be included. We further expand our model for clamping losses in Sec.~\ref{sec:Clamping} and intrinsic losses in Sec.~\ref{sec:Intrinsic}.

\subsection{Clamping Loss}\label{sec:Clamping}

\begin{figure}[h!]
\includegraphics[width=\columnwidth]{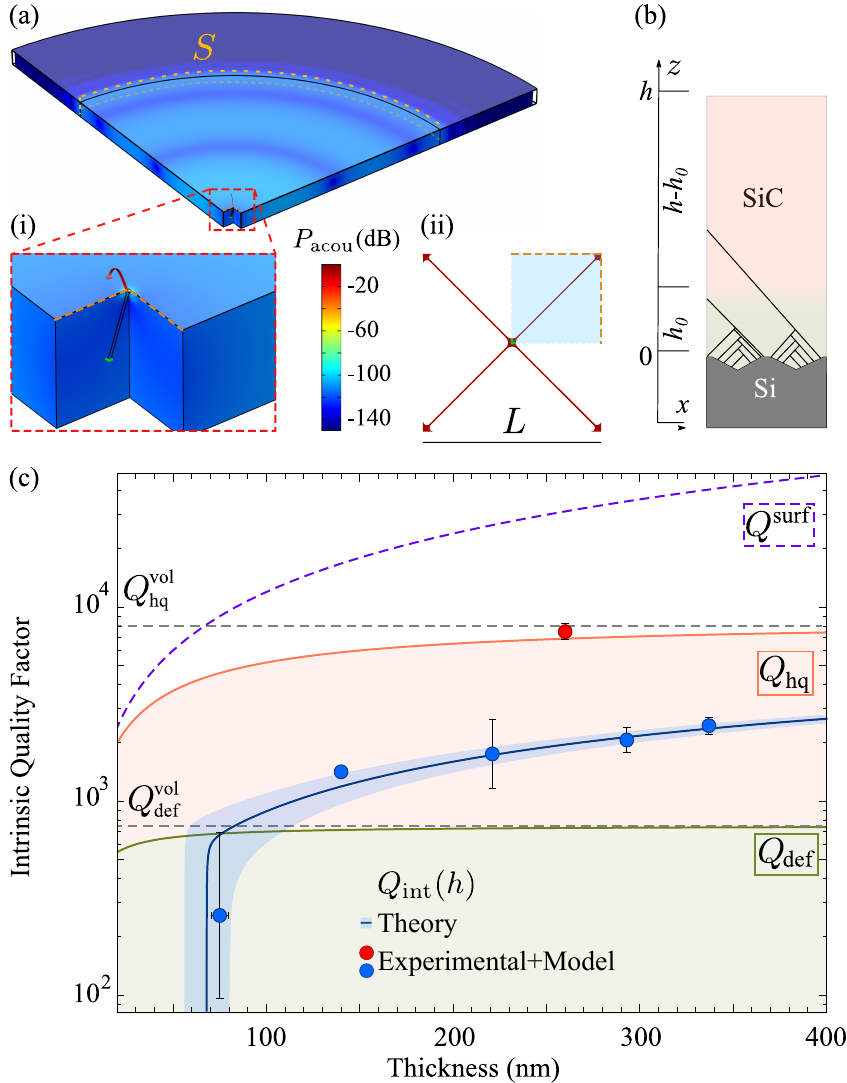}
\caption{(a-i) FEM simulation of the acoustic power $P_{\textnormal{acou}}$ propagating through the Si substrate, with perfectly matched layer at the surface $S$, calculated in a quarter of the total domain using symmetries. Zoom in of the quarter of the trampoline used for the calculation. (a-ii) Top-view of the trampoline design where the blue shaded region represents the domain used during the FEM simulation. The clamping boundaries with color lines are mapped into the substrate for reference. (b) Schematic of the lateral view of the SiC single crystal atop Si, based on TEM studies and depictions~\cite{iacopi_evidence_2013,sun_saddle-shape_2012}. The region $h_0$ represents the thickness of the defect-rich layer localized near the interface. The lines near the interface illustrate stacking defects and dislocations. (c) Thickness dependent fitting parameters of the bi-layer system. Intrinsic $Q_{\textnormal{def}}$ of the defect-rich layer (green line) and the high-quality single crystal $Q_{\textnormal{hq}}$ (orange line). Thickness dependent surface loss $Q^{\textnormal{surf}}(h)$ (purple dashed line) of the bi-layer system. Each layer is limited by their respective volume losses $Q_{\textnormal{vol}}^{\textnormal{hq,def}}$ (gray dashed lines). Theoretical fit for $Q_{\textnormal{int}}$ from Eq.~\eqref{eq:Intrinsic_Bilayer} for the front-side etched resonators compared to experimental data, with the blue-shaded region being the uncertainty of the fit.}\label{fig:Intrinsic}
\end{figure}

Clamping losses are caused by phonons tunneling from the resonator into the substrate~\cite{cole_phonon-tunnelling_2011}. The elastic energy leaks out of the resonator through the clamping points in the form of acoustic radiation. The amount of leakage depends on the impedance mismatch between the resonator and the substrate~\cite{schmid_fundamentals_2016}. To estimate the clamping losses produced in our system, we use finite element modelling (FEM) to calculate the total elastic energy $U$ stored in the resonator and the power $P_{\textnormal{acou}}$ carried by the acoustic radiation. The clamping loss dominated quality factor $Q_{\textnormal{clamp}}$ can be estimated as the ratio of energy stored versus energy lost during one oscillation cycle as ${Q_{\textnormal{clamp}}
=
2\pi \omega_{\textnormal{m}}} 
\frac{U}{\langle P_{\textnormal{acou}} \rangle}$. In Fig.~\ref{fig:Intrinsic}(a-i), we show the calculation of $P_{\textnormal{acou}}$ for a trampoline attached to a substrate of thickness $h_{\textnormal{s}}=500~\upmu$m, where symmetries are used to reduce the computational demand calculating over one quarter of the domain. The trampoline design is represented in Fig.~\ref{fig:Intrinsic}(a-ii), where the blue shaded region is the quarter of the domain used during the calculations. The power crossing the interface $S$ leaves the substrate where a perfectly matched layer attenuates it and prevents it from reflecting. The parameters used in the FEM calculation included the Young's modulus of the Si substrate $E_\textnormal{s}=170~$GPa, and the density $\rho_\textnormal{s}=2650$~kg/m$^3$~\cite{kermany_microresonators_2014}. 

The numerical results obtained from the FEM simulations for $Q_{\textnormal{clamp}}$ require substantial computational time and fail to predict the clamping losses at small thickness, where an analytic model represents an advantage. Trampoline resonators are two-dimensional structures that share similarities with membrane resonators, for which the analytic expression
\begin{equation}\label{eq:clamping}
Q_{\textnormal{clamp}}
= \alpha
\frac{3\rho_{\textnormal{s}}}{2 \rho} 
\sqrt{ \frac{E_{\textnormal{s}}\rho}{2\sigma(h) \rho_{\textnormal{s}}}}
\frac{L}{h},
\end{equation}
exists for the fundamental mode. Comparing our numerical solution to this analytic expression, we find an agreement for the prefactor $\alpha= 200$, which is a fitting parameter correcting for substrate imperfections and mounting conditions~\cite{villanueva_evidence_2014}. Consequently, the analytic model is used henceforth. With $\alpha$ fitted, the results obtained from FEM simulations and analytic expressions agree within 0.01~\% for $h \geq 200$~nm and within $\sim 5$\% for $h \sim 100$~nm, diverging for thinner thickness due to limitations on the meshing of the domain of the FEM simulation. In Fig.~\ref{fig:Results} we plot $Q_{\textnormal{clamp}}$ (purple dashed line) as a function of thickness showing it is not the primary limitation of the performance of our resonators, but cannot be entirely neglected.

\subsection{Intrinsic Dissipation}\label{sec:Intrinsic}
Intrinsic dissipation ($Q_{\textnormal{int}}^{-1}$) in micro-resonators originates from two main sources, surface ($Q_{\textnormal{surf}}^{-1}$) and volume losses ($Q_{\textnormal{vol}}^{-1}$). The dominant contribution to surface loss is expected to occur on the top and the bottom surfaces of the device, since these have much larger area compared to lateral surfaces~\cite{villanueva_evidence_2014}. Volume losses in crystalline resonators are caused by defect motion within the resonator volume~\cite{cleland_noise_2002}. In materials with a non-uniform vertical density of defects, such as hetero-epitaxially grown 3C-SiC, the dissipation profile is expected to be vertically non-uniform.

As a simple model of the non-uniform distribution of defects we consider the SiC film as a bi-layer system. The first layer is a defect-rich layer (def) near the Si-SiC interface with thickness $h_0$, as represented in Fig.~\ref{fig:Intrinsic}(b). The second layer is a high-quality single crystal (hq) with thickness $h-h_0$, above the defect-rich layer expected to have a significantly reduced but not eliminated density of defects (black lines)~\cite{iacopi_orientation-dependent_2013}. In order to identify the different contributions to intrinsic dissipation in this bi-layer system, two main assumptions are made: first that each layer has independent volume and surface dissipation components; and second, that the surface dissipation follows an inverse linear relation with thickness, as has been shown for silicon nitride membranes and microcantilevers~\cite{villanueva_evidence_2014,
yasumura_quality_2000}. For the high quality crystal ($h>h_0$) the surface loss is then given by ${Q_{\textnormal{hq}}^{\textnormal{surf}}(h)=\beta_{\textnormal{hq}} (h-h_0)}$ for the high quality layer, and ${Q_{\textnormal{def}}^{\textnormal{surf}}=\beta_{\textnormal{def}}\times h_0}$ for the defect-rich layer. 

The total intrinsic dissipation $Q_{\textnormal{int}}^{-1}(h)$ is given by the weighted sum of the dissipation of each layer  
\begin{equation}\label{eq:Intrinsic_Bilayer}
\begin{array}{ll}
Q_{\textnormal{int}}^{-1}(h)
=&
\displaystyle\left(\frac{h-h_0}{h}\right)
\left[
(Q_{\textnormal{hq}}^{\textnormal{surf}}(h))^{-1}
+
(Q_{\textnormal{hq}}^{\textnormal{vol}})^{-1}
\right]\\
\\
& +
\displaystyle\left(\frac{h_0}{h}\right)
\left[
(Q_{\textnormal{def}}^{\textnormal{surf}})^{-1}
+
(Q_{\textnormal{def}}^{\textnormal{vol}})^{-1}
\right],
\end{array}
\end{equation}
where $Q_{\textnormal{hq}}^{\textnormal{vol}}$ and $Q_{\textnormal{def}}^{\textnormal{vol}}$ are the volume quality factors of the high quality crystal and defect-rich layer, respectively. The experimentally determined values of $Q_{\textnormal{int}}$ (blue dots) shown in Fig.~\ref{fig:Intrinsic}(c) are deduced from Eq.~\eqref{eq:Q}, by combining the experimental $Q$ presented in Fig.~\ref{fig:Results} with the analytic models for $Q_{\textnormal{clamp}}$ and $\mathcal{D}(h)$. Meanwhile, the theory fit is obtained from simultaneously fitting the five parameters of the bi-layer SiC model $Q_{\textnormal{def}}^{\textnormal{vol}}$, $Q_{\textnormal{hq}}^{\textnormal{vol}}$, $\beta_{\textnormal{hq}}$, $\beta_{\textnormal{def}}$ and $\alpha$, to the six data points in Fig.~\ref{eq:Intrinsic_Bilayer}(c). The total surface loss in the bi-layer model is given by ${Q^{\textnormal{surf}}(h)= \left[\left(\frac{h-h_0}{h}\right)(Q_{\textnormal{hq}}^{\textnormal{surf}}(h))^{-1}+\left(\frac{h_0}{h}\right)(Q_{\textnormal{def}}^{\textnormal{surf}})^{-1}\right]^{-1}},$
and is shown as a function of film thickness (blue dashed line). The fitting parameters are summarized in Table~\ref{table:Qs}.\begin{table}[h!]
\begin{tabular}{ |c|r l|  }
 \hline\rule{0pt}{12pt}
Parameter & Value & \\
 \hline\rule{0pt}{15pt}
$Q_{\textnormal{def}}^{\textnormal{vol}}$& (0.75$\pm$0.15)& $\times 10^3$ \\ \rule{0pt}{15pt}
$Q_{\textnormal{def}}^{\textnormal{surf}}$ & (10$\pm$ 4)& $\times 10^{10}~\textnormal{m}^{-1}\times h_0$\\ \rule{0pt}{15pt}
$Q_{\textnormal{hq}}^{\textnormal{vol}}$ & (8.0$\pm$1.8)& $\times 10^3$\\ \rule{0pt}{15pt}
$Q_{\textnormal{hq}}^{\textnormal{surf}}$ &(12$\pm$ 5)& $\times 10^{10}~\textnormal{m}^{-1}\times (h-h_0)$ \\ \rule{0pt}{15pt}
$h_0$ & (68$\pm$12)&$\times 10^{-9}$~m\\ 
\hline
\end{tabular}
\caption{Volume and surface quality factors of the defect-rich layer and the high-quality layer.}\label{table:Qs}
\end{table}

Volumetric dissipation is associated to friction among crystal dislocations, stacking faults, or defects in the bulk SiC. Accordingly, the dissipation is expected to be higher in regions of the film with high defect density than in regions with low defect density. Our fitting parameters ${Q_{\textnormal{hq}}^{\textnormal{vol}}=(8.0\pm 1.8) \times 10^{3}}$ and ${Q_{\textnormal{def}}^{\textnormal{vol}}=750\pm 150}$ are depicted as gray dashed lines in Fig.~\ref{fig:Intrinsic}(c), where the green and orange lines illustrate the intrinsic quality factor for trampolines made solely from the defect-rich layer and high quality layer respectively. The fact that $Q_{\textnormal{hq}}^{\textnormal{vol}}$ is more than an order of magnitude higher than $Q_{\textnormal{def}}^{\textnormal{vol}}$
confirms that the defect layer does indeed have significantly degraded quality factor due to crystalline interface defects and these defects increase intrinsic dissipation. 

The surface loss of the defect and high-quality layers, ${\beta_{\textnormal{def}}=(10\pm 4)\times 10^{10}~\textnormal{m}^{-1}}$ and ${\beta_{\textnormal{hq}}=(12 \pm 5)\times 10^{10}~\textnormal{m}^{-1}}$, respectively, are identical within the uncertainty of the fit. This suggests that surface losses are not drastically affected by the etching process, and that interfacial defects have an impact on the volume component of the dissipation rather than the surface component; consistent with the chemical stability of the surface composition of SiC~\cite{mehregany_silicon_1998,azevedo_sic_2007}.

The highest quality factor of $Q=1.74\times 10^6$ was obtained on a back-side etched resonator. The average $Q$ of the back-side etched resonators is about five times higher than the predicted average value for front-side etched resonators of the same thickness, and more than an order of magnitude higher than the $Q$ measured of the thinnest front-side etched resonators. This results from an almost one order of magnitude increase in the intrinsic quality factor $Q_{\textnormal{int}}$ of back-side etched resonators is enhanced by almost an order of magnitude compared to front-side etched resonators of similar thickness. The $Q$ of front- and back-side etched resonators are limited primarily by their corresponding volume losses. 

While already comparable with previous crystalline resonators~\cite{onomitsu_ultrahigh-_2013,liu_high-q_2011,cole_tensile-strained_2014}, our results show that SiC resonators have significant possibilities for further improvement. Even though the high-quality single crystal layer has a low density of defects, these are not eliminated~\cite{iacopi_orientation-dependent_2013}. The volumetric part of the intrinsic quality factor, while remains well below the single crystal limit for SiC $Q_{\textnormal{hq}}^{\textnormal{vol}}\sim 10^{5}$~\cite{ashby_overview_1989} positively compares to LPCVD silicon nitride resonators that have already reached the volume loss limit for their amorphous composition~\cite{villanueva_evidence_2014,sadeghi_influence_2019,ghadimi_radiation_2017}. The complete removal of interfacial crystal defects in SiC could therefore potentially lead to exceptional enhancements of the quality factor as high as $\mathcal{D}(h)\times Q_{\textnormal{int}}\sim 10^8$ for a trampoline of thickness $h=70$~nm. Moreover, implementing dissipation engineered designs exploiting the high yield strength of SiC could allow quality factors as high as $Q\sim 6\times 10^9$~\cite{ghadimi_elastic_2018,fedorov_generalized_2019}.

\section{Conclusion}

This paper has explored the possibility of achieving ultrahigh quality factors in crystalline thin-film microresonators. Our results verify the prediction that interfacial defects within the thin films can severely degrade the intrinsic quality factor of crystalline resonators. We develop a crystalline-material-compatible back-side etch procedure to remove this layer, which can be applied to enhance the instrinsic quality factor of other crystalline stressed resonators made from epitaxially grown materials~\cite{onomitsu_ultrahigh-_2013, buckle_stress_2018,
liu_high-q_2011,
cole_tensile-strained_2014}.

Our method allows a factor of five improvement in diluted quality factor for single-crystal silicon carbide resonators, by increasing the intrinsic quality factor, achieving values of $Q>10^6$. By developing a detailed model of the dissipation in bi-layer films, we are able to precisely determine the material properties of SiC epitaxial films with higher resolution than spectroscopic techniques. Our model predicts that diluted quality factors as high as $6 \times10^9$ may be possible using both dissipation and strain engineering if high quality single-crystal silicon carbide was used. \\

\section*{Acknowledgement}

This research was funded by the Australian Research Council and Lockheed Martin Corporation through the Australian Research Council Linkage Grant LP140100595. This research is partially supported by the Commonwealth of Australia as represented by the Defence Science and Technology Group of the Department of Defence. Support was also provided by the Australian Research Council Centre of Research Excellence for Engineered Quantum Systems (CE110001013). E. R. and V. M. V acknowledge CONACYT (381542 and 234733 respectively). V. M. V acknowledges LN-293471 and LN-299057. W.P.B. acknowledges a Future Fellowship from the Australian Research Council (FT140100650). The 3C-SiC material was developed and supplied by Leonie Hold and Alan Iacopi of the Queensland Microtechnology Facility (QMF), within the Queensland node of the Australian National Fabrication Facility. This work was performed in part at the Queensland node of the Australian National Fabrication Facility (ANFF), a company established under the National Collaborative Research Infrastructure Strategy to provide nano and microfabrication facilities for Australia’s researchers. The authors acknowledge James Bennett for useful discussions and technical support.

\bibliographystyle{apsrev4-1}
%
\end{document}